\documentclass[9pt,twocolumn,twoside]{pnas-new}
\templatetype{pnasresearcharticle} % Choose template 
\title{Dynamic Gardner crossover in a simple glass}

\author[a,b]{Qinyi Liao}
\author[c,d]{Ludovic Berthier} 
\author[a,e]{Hai-Jun Zhou}
\author[b,f]{Ning Xu}

\affil[a]{CAS Key Laboratory for Theoretical Physics, Institute of Theoretical Physics, Chinese Academy of Sciences, Beijing 100190, China}
\affil[b]{Department of Physics, University of Science and Technology of China, Hefei 230026, P. R. China}
\affil[c]{Laboratoire Charles Coulomb (L2C), University of Montpellier, CNRS, 34095 Montpellier, France}
\affil[d]{Yusuf Hamied Department of Chemistry, University of Cambridge, Lensfield Road, Cambridge CB2 1EW, United Kingdom}
\affil[e]{MinJiang Collaborative Center for Theoretical Physics, MinJiang University, Fuzhou 350108, China}
\affil[f]{Hefei National Research Center for Physical Sciences at the Microscale, CAS Key Laboratory of Microscale Magnetic Resonance, Hefei 230026, P. R. China}

\leadauthor{Liao} 

\significancestatement{The jamming transition is crucial to understand the formation and the nature of ubiquitous amorphous solids. Only recently, jamming has been associated with a thermodynamic Gardner phase of glasses by a replica mean-field theory. However, many realistic amorphous solids are formed far from equilibrium, and the generality of Gardner physics remains elusive. We address this question by simulating a simple glass model of hard disks. We discuss the existence of a dynamic Gardner crossover for anomalous dynamics and provide the complete glass phase diagram by analysing the landscape complexity. Our results bridge off-equilibrium experiments and thermodynamic studies and pave the way for constructing a dynamical glass theory that includes marginally stable phases.}

\authorcontributions{Q.L., L.B., H.Z., and N.X. designed the project. Q.L. performed research. All the authors analysed the data and wrote the paper.}
\authordeclaration{The authors declare no competing interests.}
\correspondingauthor{\textsuperscript{2}To whom correspondence should be addressed. E-mail: qinyi.liao.phy@gmail.com or ludovic.berthier@umontpellier.fr or zhouhj@itp.ac.cn or ningxu@ustc.edu.cn}


\begin{abstract}
The criticality of the jamming transition responsible for amorphous solidification has been theoretically linked to the marginal stability of a thermodynamic Gardner phase. While the critical exponents of jamming appear independent of the preparation history, the pertinence of Gardner physics far from equilibrium is an open question. To fill this gap, we numerically study the nonequilibrium dynamics of hard disks compressed towards the jamming transition using a broad variety of protocols. We show that dynamic signatures of Gardner physics can be disentangled from the aging relaxation dynamics. We thus define a generic dynamic Gardner crossover regardless of the history. Our results show that the jamming transition is always accessed by exploring increasingly complex landscape, resulting in anomalous microscopic relaxation dynamics that remains to be understood theoretically.
\end{abstract}

\dates{This manuscript was compiled on \today}
\doi{\url{www.pnas.org/cgi/doi/10.1073/pnas.XXXXXXXXXX}}

\begin{document}

\maketitle
\thispagestyle{firststyle}
\ifthenelse{\boolean{shortarticle}}{\ifthenelse{\boolean{singlecolumn}}{\abscontentformatted}{\abscontent}}{}

%% If your first paragraph (i.e. with the \dropcap) contains a list environment (quote, quotation, theorem, definition, enumerate, itemize...), the line after the list may have some extra indentation. If this is the case, add \parshape=0 to the end of the list environment.
%\dropcap{T}his PNAS journal template is provided to help you write your work in the correct journal format. Instructions for use are provided below. 

%Note: please start your introduction without including the word ``Introduction'' as a section heading (except for math articles in the Physical Sciences section); this heading is implied in the first paragraphs. 

\dropcap{T}he jamming transition describes the formation of amorphous solids in materials composed of repulsive particles~\cite{LN98, OSLN03}. It has been the subject of important research activity in the last decades, encompassing statistical mechanics analysis, numerical studies, and experimental investigations of granular and colloidal materials~\cite{liu2010jamming,van2009jamming}. The jamming transition also attracts interest in contexts such as the geometry of sphere packings, the statistical mechanics of liquid and glass states, and is related to mechanical properties of biophysical matter and machine learning in computer science~\cite{BM62,DTS05,xu2010anharmonic,bi2016motility,geiger2019jamming}. Jammed packings have original physical properties that differ dramatically from crystalline solids. In particular, the application of small perturbations often leads to large-scale responses, showing that jammed materials are marginally stable~\cite{MW15}.

Understanding the structure of jammed packings is difficult because the absence of thermal fluctuations prevents the use of a statistical ensemble, as first noted by Edwards~\cite{EO89,henkes2009statistical}. This key problem was circumvented by describing jamming as the end-point of the compression of dense assemblies of thermalised soft or hard repulsive spheres~\cite{parisi2020theory}. This approach is meaningful because Brownian particles near jamming belong to dynamically arrested glass states~\cite{berthier2009glass}. It becomes possible to follow quasi-statically the evolution of glassy states compressed towards jamming which only probes a restricted region of configuration space~\cite{KZ10,RUYZ14}. For this thermodynamic description to be correct, it is crucial that the compression history occurs over timescales that are much shorter than the timescales related to the crossing of barriers in configuration space. In the limit of large dimensions, $d \to \infty$, where the analytic calculations were performed, this timescale separation is guaranteed by the divergence of free energy barriers separating distinct glass states~\cite{parisi2020theory}.

An unexpected outcome was the discovery that glass states followed quasi-statically upon compression undergo a Gardner phase transition separating two types of glass phases described by distinct sets of solutions~\cite{CKPUZ13,berthier2019gardner}. The Gardner phase bears many similarities with the replica symmetry broken spin glass phase found at low temperatures in models of disordered magnets~\cite{Ga85, GKS85}, which implies the existence of specific aging effects with multiple timescales and lengthscales and a hierarchical free energy landscape~\cite{CKPUZ14}. The Gardner phase directly impacts the physical properties of jammed materials, as shown by mean-field studies~\cite{KPUZ13,yoshino2014shear, RUYZ14, FPUZ15}. More broadly, Gardner physics arises from breaking the replica symmetry and its study is of great interest in fields encompassing optimization problems~\cite{montanari2021optimization,alaoui2020algorithmic}, deep learning~\cite{yoshino2020complex}, cell tissues~\cite{pinto2022hierarchical}, magnetic materials~\cite{kumar2020study}, polydisperse crystals~\cite{charbonneau2019glassy, kool2022gardner}, etc. The dynamic Gardner crossover analysed in this work in the hard disk model might apply to other complex systems with rough landscapes, as studied by these different communities.

It is not known whether the Gardner transition can survive in physical dimensions~\cite{urbani2015gardner,charbonneau2017nontrivial}. Yet the measured critical exponents of jamming are independent of the dimension for $d \geq 2$~\cite{OSLN03}. The robustness of jamming criticality is difficult to reconcile with the fragility of the Gardner transition. More questionable is the fact that jamming was studied using different approaches which cannot always be described using quasi-static thermodynamic descriptions~\cite{OSLN03,berthier2009glass}. There is therefore a deep theoretical gap between conventional jamming studies and the available thermodynamic approach. Even within mean-field spin glasses, the off-equilibrium dynamics of glasses entering the Gardner phase has not been studied beyond some incomplete attempts~\cite{barrat1997temperature, marinari2000off, MR03, rizzo2013replica, altieri2020dynamical}. In fact, there are increasing efforts to tackle off-equilibrium glassy dynamics~\cite{manacorda2020numerical,verstraten2021time,sibani2021record,lubchenko2017aging,oyama2021shear}, but the dynamic Gardner crossover itself remains out of reach of those attempts, and appears as a very difficult task. Our results may help shed light on the expected physicsal behaviour.

Numerical simulations and experiments were performed to detect signs of the Gardner transition~\cite{BCJPSZ15, SD16, SZ18, xiao2021probing, PhysRevLett.126.028001,wang2022experimental}, or to explore the physical consequences of the hierarchical free energy landscape for the aging dynamics~\cite{LB18,scalliet2019nature,artiaco2020exploratory} and the mechanical properties~\cite{JUZY18,wang2022experimental} of dense assemblies of repulsive particles. Although the Gardner transition is inexistent in $d=2$, the physics of hard disks is reminiscent of its $d=3$ counterpart with clear signatures of a strong Gardner crossover~\cite{LB18}. Many numerical studies mimicked the analytic thermodynamic construction using highly stable configurations to reproduce the analytic state-following construction~\cite{BCJPSZ15, SZ18, LB18,scalliet2019nature, artiaco2020exploratory, JUZY18}. Very few studies~\cite{SZ18,charbonneau2021memory} attempted to characterise the Gardner phase and related microscopic dynamics for the less stable systems studied experimentally with mixed conclusions. On the other hand, despite recent reports on Gardner-like phenomena in the systems of long-ranged interactions~\cite{shang2020elastic,oyama2021shear}, the relevance of Gardner physics in soft glasses is debated~\cite{PhysRevLett.120.085705, hammond2020experimental,scalliet2019nature, PhysRevLett.126.028001}. Our work  is helpful to resolve the controversy and understand the stability of those glassy materials.

There is a growing interest in Gardner-related phenomena in a wide range of complex systems but results are not always conclusive because the studied models and protocols are often very complex. Using well-controlled methods and well-studied systems is useful. Hard disks represent a simple and practical choice with quite a venerable history across statistical physics, which represents a fruitful playground to study Gardner physics beyond mean-field theory. Here, we simulate a two-dimensional system of hard disks~\cite{LB18} and use a wide range of preparation protocols from poorly annealed systems to stable ones to explore the nonequilibrium dynamics observed during compression protocols towards the jamming transition. We find that the signs of a Gardner crossover are robust even in the regime where strong structural relaxations are present and can be disentangled from the aging relaxation dynamics. We thus generalize the thermodynamic Gardner crossover to a dynamic crossover for any specific history, as well as propose a complete phase diagram for the emergence of Gardner physics for hard disks, thus going much beyond the purely static approach followed in Ref.~\cite{LB18}. Our results show that Gardner physics is at play in the approach to jamming, independently of the preparation history.

\section*{Results}

\subsection*{Sample preparation}

%Include department, institution, and complete address, with the ZIP/postal code, for each author. Use lower case letters to match authors with institutions, as shown in the example. PNAS strongly encourages authors to supply an \href{https://orcid.org/}{ORCID identifier} for each author. Individual authors must link their ORCID account to their PNAS account at \href{http://www.pnascentral.org/}{www.pnascentral.org}. For proper authentication, authors must provide their ORCID at submission and are not permitted to add ORCIDs on proofs.

We simulate dense assemblies of two-dimensional hard disks~\cite{LB18} with a continuous size polydispersity to avoid crystallization (see Methods). The mean disk diameter sets the unit length, and the pressure is normalised by $k_{B}T \rho$, with $\rho$ the number density and $k_{B}T$ the thermal energy defining the normalised pressure $Z$. Since the scale set by the potential energy is infinite for hard disks, we can impose $k_{B}T = 1$. The simulations are performed using a Monte Carlo (MC) algorithm, and the MC sweep is used as the time unit. The number of particles is $N = 1024$, chosen to allow exploration of the free energy landscape~\cite{LB18}. We remove contributions from the translational Mermin-Wagner fluctuations to the dynamics using cage-relative coordinates when computing physical observables (see Methods)~\cite{mw2017,shiba2016unveiling,vivek2017long}.

To search for Gardner physics for the whole glass phase diagram, we devise a two-step sample preparation. The prepared samples will then be employed as starting points of compression protocols devised to probe the underlying free energy landscape. Let us first introduce the preparation. We start by equilibrating the system at fixed volume fractions $\varphi_g$ using constant volume MC simulations. To achieve equilibrium for dense configurations, we use the swap MC algorithm to speed up the thermalization~\cite{berthier2016equilibrium}. This enables us to access equilibrium fluid states at densities much higher than the dynamic mode-coupling theory (MCT) crossover, $\varphi_{MCT} \approx 0.795$. In the second step, we switch off the swap MC moves and perform conventional MC simulations in the NPT ensemble at various reduced pressure $Z_s \geq Z(\varphi_g)$. Constant pressure MC simulations are convenient to approach the jamming transition which is obtained in the limit $Z \to \infty$. We let the system relax at pressure $Z_s$ for a time $t_s$ and store the final configurations, which are characterised by three control parameters $(\varphi_g, Z_s, t_s)$. At the end of this preparation, the configurations are off-equilibrium at state point $(Z_s, \varphi_s)$. We repeat this recipe starting from independent configurations at $\phi_g$ when performing ensemble averages over independent trajectories. 

Therefore, this two-step protocol enables us to cover multiple state points and physical regimes, in particular those characterised by aging relaxation dynamics far from equilibrium, which could be related to a large class of experimental situations~\cite{SD16, PhysRevLett.126.028001,xiao2021probing,wang2022experimental}. The samples drawn from a compression with a large $Z_s = Z(\phi_g)$ (equivalently, $t_s = 0$) correspond to previous studies~\cite{BCJPSZ15,SZ18,LB18} mimicking the theoretical state-following construction. Adding the parameter $t_s$ for lower $\phi_g$ values is an important new feature of our study which allows us to explore in a controlled manner and with minimal ingredients a broad range of nonequilibrium states with different degrees of aging and stability.

\begin{figure*}
\centering
\includegraphics[width=.95\linewidth]{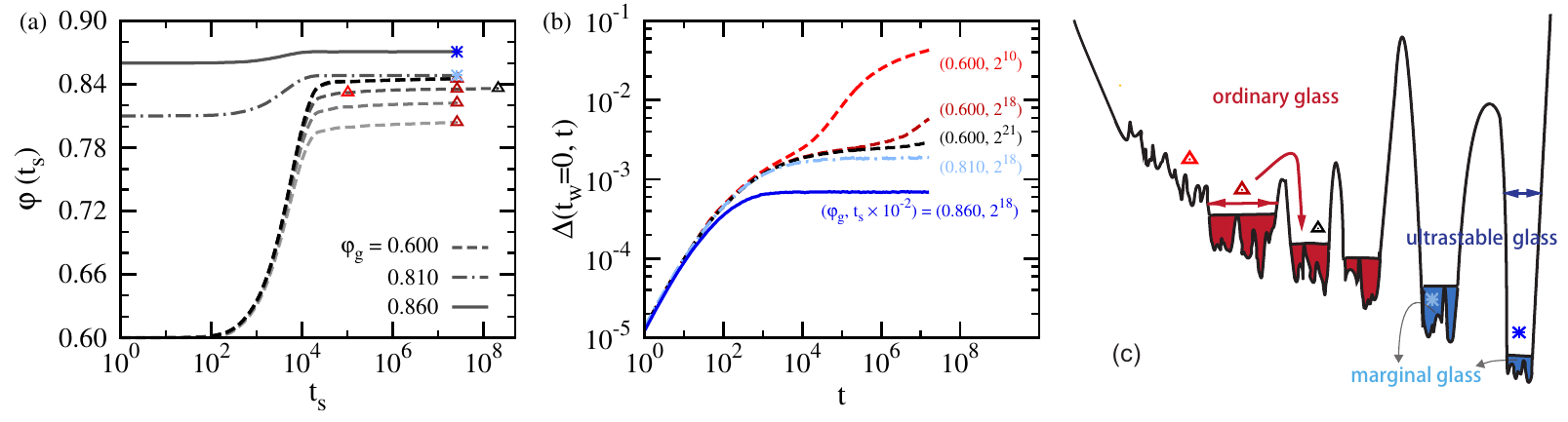}
\caption{Broad range of sample preparation. (a) Evolution of the packing fraction in the second step of the sample preparation, starting from different $\varphi_g$ at $t_s=0$ and compressing at constant pressure $Z_s$. For $\varphi_g = 0.600$ we show $Z_s = 30$, 50, $10^2$, $10^3$ and $10^4$ (from bottom to top), while for $\varphi_g = 0.810$ and $\varphi_g = 0.860$ we only show $Z_s = 10^2$. Representative glass samples characterised by $(\phi_g, Z_s, t_s)$ are marked by symbols. (b) The caged-relative mean-squared displacements $\Delta (t_w=0, t)$ for glass samples at pressure $Z_s = 10^2$ marked in (a) with $(\phi_g, t_s)$ as indicated. Strong aging relaxation effects are obvious for ordinary glasses prepared from $\phi_g=0.600$. (c) Sketch of the position of the different types of glasses shown in (a) in the free energy landscape, where the height encodes the degree of stability and the black horizontal lines represent the Gardner crossover.}
\label{fig:fig1} 
\end{figure*}

We display several examples of the NPT second stage of the preparation in Fig.~\ref{fig:fig1}(a), where each line represents the evolution of the volume fraction $\varphi(t_s)$ for pairs of $(\varphi_g, Z_s)$ values. The different symbols indicate representative preparation states characterised by $(\phi_g, Z_s, t_s)$ that will be used to investigate the Gardner crossover. The pressure $Z_s$ for states initialized from ergodic liquids with $\varphi_g = 0.600$ ranges from $Z_s=30$ to $Z_s=10^4$. When starting from denser glass states with $\varphi_g = 0.810$ and $\varphi_g = 0.860$, we only show results for $Z_s = 10^2$. When compressing from $\varphi_g = 0.600 \ll \varphi_{MCT} \approx 0.795$, it can be seen that $\varphi$ grows with the time $t_s$ over the entire window, showing that $t_s$ is a relevant control parameter for such nonequilibrium preparation histories. Glasses obtained after longer $t_s$ are more stable and, accordingly, display smaller particle diffusion during the densification, see Fig.~\ref{fig:fig1}(b). The mean-squared displacements reveal the presence of aging dynamics for these systems. We shall call these glassy states ``ordinary glasses''.

The situation is qualitatively different for the case of $(\varphi_g = 0.860, Z_s = 10^{2})$, where the system rapidly achieves restricted equilibrium inside the metastable glass basin in a time shorter than $t_s \approx 10^{5}$. The mean-squared displacement for that system in Fig.~\ref{fig:fig1}(b) displays a stable plateau. We refer to these glasses as ``ultrastable glasses'', and they are roughly similar to systems used in earlier numerical studies \cite{BCJPSZ15,SZ18,LB18}. Finally, glasses prepared by compressing from $\varphi_g = 0.810$ to $Z_s = 10^2$ are close to the thermodynamic Gardner crossover that is estimated below to occur near $Z_G \approx 10^2$. They show a mild time dependence of $\varphi(t_s)$ and a larger mean-squared displacement. We call these glasses ``marginal glasses''. For such large packing fractions, the physical MC dynamics is nearly arrested.  

In Fig.~\ref{fig:fig1}(c), we offer a sketch of the position in the energy landscape of the three categories of glasses prepared in Fig.~\ref{fig:fig1}(a), namely ordinary, marginal and ultrastable glasses. We are particularly interested in studying the pertinence of the Gardner physics for ordinary glasses which are not well understood theoretically and are closer to systems studied experimentally. To this end, we will compare their study with results obtained on ultrastable and marginal glasses. 

\subsection*{Exploring complex landscapes upon compression}

%All authors must submit their articles at \href{http://www.pnascentral.org/cgi-bin/main.plex}{PNAScentral}. If you are using Overleaf to write your article, you can use the ``Submit to PNAS'' option in the top bar of the editor window. 

After the two-step sample preparation described above, we perform various compression histories using NPT MC simulations. The end of the sample preparation at $t_s$ corresponds to the waiting time $t_w=0$, after which the pressure is instantaneously changed to a new value $Z > Z_s$, or kept at $Z=Z_s$ as in Fig.~\ref{fig:fig1}(b). Although this may superficially resemble the state-following construction proposed by thermodynamic theories, our numerical approach can be applied to all type of glass states and we do not assume that infinitely long-lived states exist.

We instantaneously change the pressure from the preparation pressure $Z_s$ to a range of applied pressure $Z$ at time $t_w=0$ and track the evolution of the system as a function of the time $t_w$ spent at the final pressure $Z$. Following earlier work, we simulate $N_c$ independent trajectories, or clones, for each prepared sample at $t_w=0$ by using different sequences of random numbers in the MC simulations. We further average over $N_s$ independent samples for each state characterised by $(\varphi_g, Z_s, t_s)$ to average over the disorder, which results in $N_s \times N_c$ simulations in total to fully characterise a given preparation history. We use $(N_s = 100, N_c = 10)$ to get statistical properties, and increase the number of clones to $N_c = 100$ when analyzing more finely the landscape structure of individual samples.

It is useful to compare the results obtained for ordinary glasses with more stable glasses. To this end we select three examples prepared at the same values $(Z_s = 200, t_s = 2^{18}\times 100)$ but for different initial equilibrium states of $\phi_g = 0.600$, $0.820$ and $0.860$ which respectively represent typical ordinary, marginal and ultrastable glasses. The pressure $Z$ varies from $Z=200$ to $Z=10^4$. 

We study the dynamics using the cage-relative mean-squared displacement (MSD) $\Delta(t_w, t)$ between times $t_w$ and $t_w + t$ at a series of waiting times $t_w$. For ultrastable glasses with $(\varphi_g = 0.860, Z_s = 200)$, the dynamics exhibits a crossover from simple vibrations to anomalous aging at a threshold pressure that coincides with the Gardner crossover $Z_{G} \approx 10^3$ (see Supplementary Fig.~\ref{app:appultra}). By contrast, we observe an aging behavior for marginal and ordinary glasses in the entire pressure range. For the marginal glass, we have confirmed that there is no diffusion to another glass state within the simulated time by decompressing these systems to a lower pressure $Z = 10^2$ where the dynamics was purely vibrational. The aging dynamics for the marginal samples is thus entirely due to the marginal stability of the Gardner phase. Instead, ordinary glasses first age at modest $Z$ because they relax from one state to another, whereas for larger $Z$ aging is instead dominated by Gardner physics, as we establish more precisely in the rest of the paper. 

\begin{figure}
\centering
\includegraphics[width=.8\linewidth]{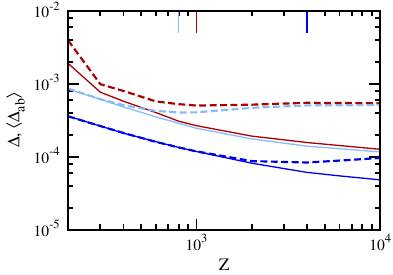}
\caption{Exploring complex landscapes upon compression. Long-time limit $(t_w = t = 2^{17}\times 100)$ of cage-relative MSD $\Delta(t_w,t)$ (full lines) and cage-relative mean-squared distance between clones $\langle \Delta_{ab}(t_w) \rangle$ (dashed lines) for three states prepared from $\phi_g = 0.600$ (red), $0.820$ (light blue) and $0.860$ (dark blue) and the same $(Z_s = 200, t_s = 2^{18}\times 100)$. The quantities $\Delta$ and $\langle \Delta_{ab} \rangle$ differ at all $Z$ for ordinary and marginal glasses, whereas they are equal up to the Gardner crossover $Z_{G}$ for ultrastable glasses. However, in all cases, $\langle \Delta_{ab} \rangle$ exhibits a nonmonotonic dependence with $Z$ as a rough landscape emerges, and the positions of the resulting minima are indicated by the vertical lines.}
\label{fig:fig2}
\end{figure}

Because of the interplay between structural relaxation and Gardner physics, it is not easy to interpret the physical origin of aging of ordinary glasses. To gain further insight, we resort to the cage-relative mean-squared distances between pairs of clones at time $t_w$, $\langle \Delta_{ab} \rangle$. In Fig.~\ref{fig:fig2}, we show the long-time values of $\Delta(t_w,t)$ and $\langle \Delta_{ab} \rangle$ measured at $t_w = t = 2^{17} \times 100$ as a function of $Z$ for the three glass states. When the landscape is simple, ergodicity is achieved within metastable basins so that the time average $\Delta$ and the ensemble average $\langle \Delta_{ab} \rangle$ coincide. Otherwise, they will differ from each other. The separation between $\Delta$ and $\Delta_{ab}$ is thus considered as the primary criterion for the emergence of a hierarchical landscape~\cite{BCJPSZ15}. This is less obvious for marginal and ordinary glasses which may undergo structural rearrangements and therefore have $\langle \Delta_{ab} \rangle$ larger than $\Delta$ throughout the entire range of pressure as observed in Fig.~\ref{fig:fig2}.

Nevertheless, the evolution of $\langle \Delta_{ab} \rangle$ with pressure itself carries interesting information about the free energy landscape. Upon compressing glasses within a hierarchical landscape, $\langle \Delta_{ab} \rangle$ grows with increasing pressure due to the rapid proliferation of states, as observed for ultrastable and marginal glasses. As shown in Fig.~\ref{fig:fig2}, $\langle \Delta_{ab} \rangle$ is nonmonotonic for all three glasses, suggesting that Gardner physics is relevant to the dynamics of ordinary glasses, despite the presence of aging relaxations.

\subsection*{Dynamic Gardner crossover}

In the state-following analysis of ultrastable glasses, the emerging aging dynamics observed beyond the Gardner crossover is related to the emergence of free energy barriers preventing the unrestricted exploration of the available phase space. As a result, when a population of clones is simulated one finds that different clones may explore different parts of this phase space, which gives rise to a broad distribution of distances between pairs of clones. Importantly, when the pressure is returned to a value smaller than the Gardner crossover, this clustering in phase space disappears and the clones evolve freely again. This represents a memory effect~\cite{LB18,scalliet2019rejuvenation}. This property is important as it enables us to distinguish aging signatures due to the Gardner crossover showing memory, from the aging relaxation towards different states, which is fully irreversible.

\begin{figure*}
\centering
\includegraphics[width=.95\linewidth]{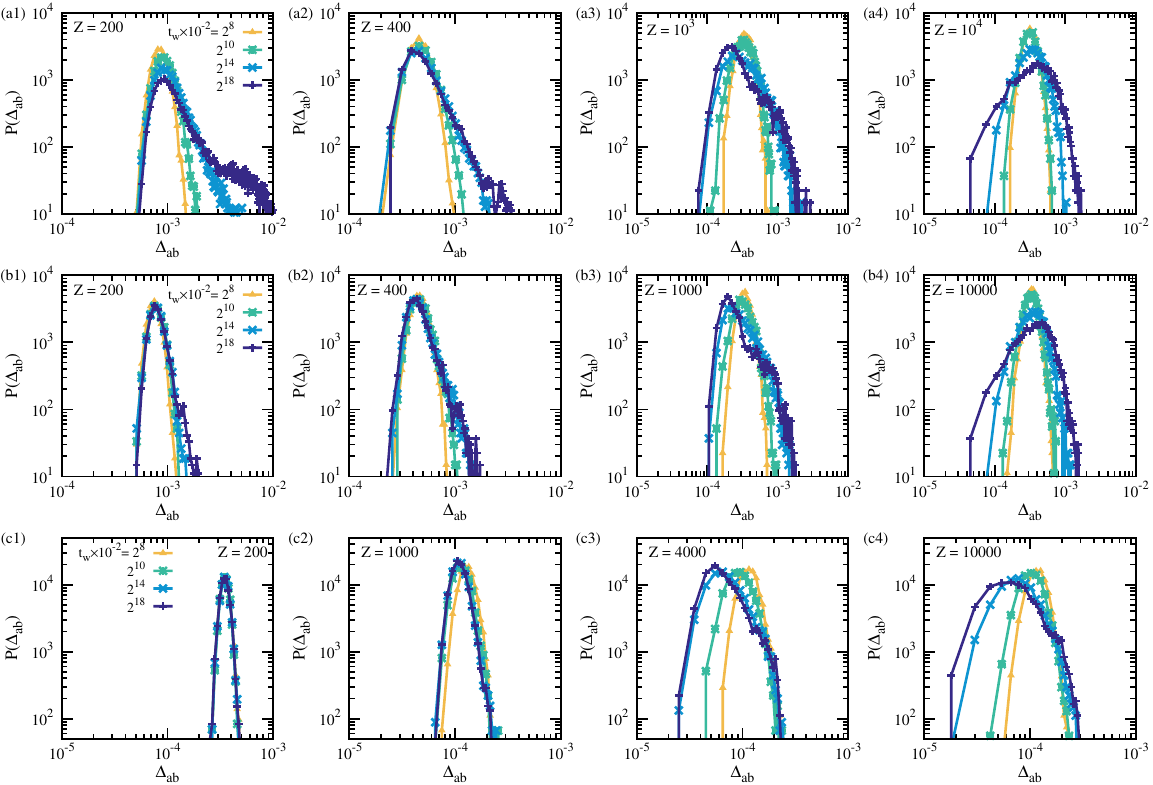}
\caption{Coexistence of aging relaxation and hierarchical landscape. 
Evolution of the distributions of cage-relative distances between clones $P(\Delta_{ab})$ at $Z = 200$ (A1-C1), $400$ (A2-C2), $10^3$ (A3-C3), $10^4$ (A4-C4) from left to right. From top to bottom, we show results for the three samples compressed from $\phi_g = 0.600$ (A1-A4), $0.820$ (B1-B4) and $0.860$ (C1-C4) to the same $(Z_s=200, t_s=2^{18}\times 100)$ as in Fig.~\ref{fig:fig2}. The distributions broaden at small $\Delta_{ab}$ when an increasingly complex landscape controls the physics at large pressure for all types of glasses. The broadening at large $\Delta_{ab}$ is pronounced at low $Z$ for less stable glasses and reflects structural relaxation.}
\label{fig:fig3}
\end{figure*}

We first plot the distributions of the cage-relative distances between clones $P(\Delta_{ab})$ in Fig.~\ref{fig:fig3} for the same three samples as in Fig.~\ref{fig:fig2}. As shown in the first column for $Z = 200$, the distribution $P(\Delta_{ab})$ for ultrastable glasses remains Gaussian at all times $t_w$, indicating unhindered vibrational motion. By contrast, the distributions $P(\Delta_{ab})$ for marginal and ordinary glasses broaden towards larger $\Delta_{ab}$ values, which corresponds to some form of aging relaxation. The influence of the initial stability is also clear, as the broadening is much stronger for the ordinary glass than it is for the marginal one and is absent for the ultrastable glass.

When the pressure is increased further, all glasses fail to achieve a complete sampling of the glass basin, and $P(\Delta_{ab})$ now develops a tail towards small values of $\Delta_{ab}$ as $t_w$ increases. It indicates that different clones explore different parts of the available phase space, some pairs being close to one another, whereas other pairs are far from each other. When the pressure is high enough, ordinary glasses behave similarly to marginal glasses, showing that Gardner physics always becomes important upon approaching jamming.

\begin{figure}
\centering
\includegraphics[width=.8\linewidth]{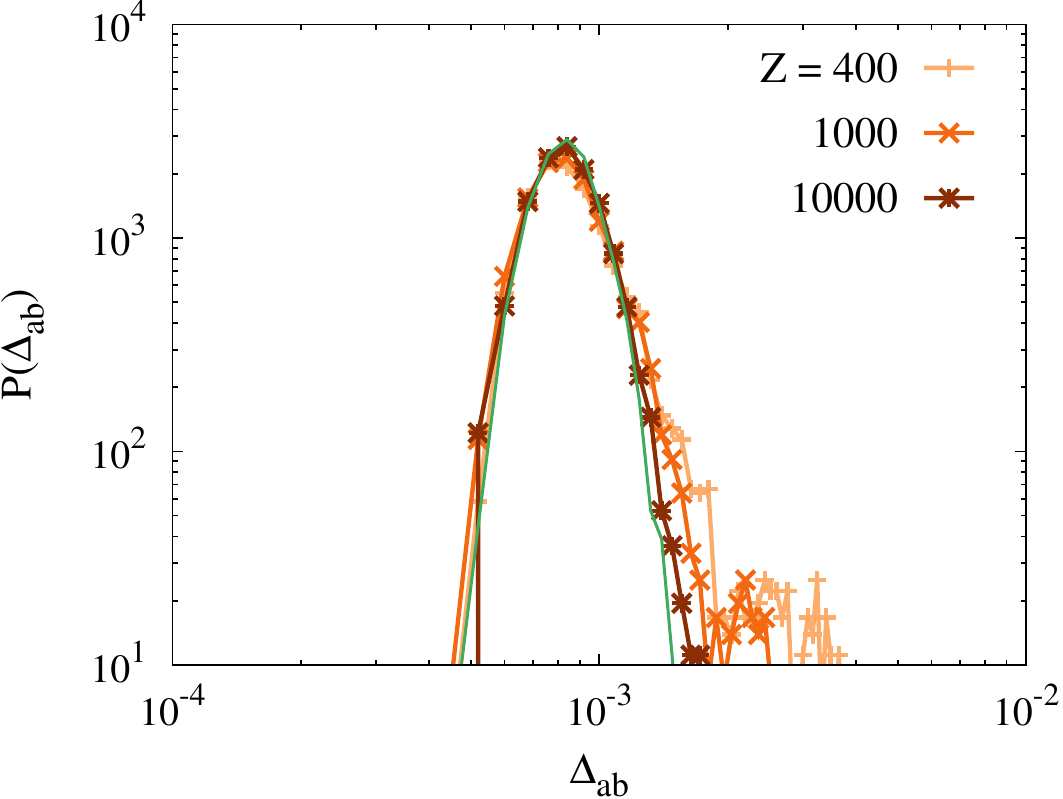}
\caption{Partial memory effect in ordinary glasses. Distributions of cage-relative distances between clones $P(\Delta_{ab})$ for ordinary glasses first compressed for $t_w = 2^{18}\times 100$ at pressure $Z$ and then decompressed back to $Z_s=200$ for a short time $t = 2^{8}\times 100$. The results are compared to the reference curve (cyan line) which only aged for a short time $t = 2^{8}\times 100$ at $Z_s=200$. Deviations with the reference distribution at large $\Delta_{ab}$ are due to aging relaxation dynamics which is irreversible.}
\label{fig:fig4}
\end{figure}

To better understand the interplay between aging dynamics due to structural relaxation or to the dynamic Gardner crossover, we decompress the clones produced from the ordinary glasses from the pressure $Z$ where they have aged for a long time back to the initial pressure $Z_s=200$ and compare the distributions to the reference case $Z=Z_s$ to investigate the presence of memory. When the non-Gaussian distributions at $Z > Z_s$ are solely due to the hierarchical Gardner landscape, the two distributions are expected to be the same. We use a short time after decompression $t = 2^{8}\times 100$ to perform this comparison as it does not allow for significant aging for the reference case, see Fig.~\ref{fig:fig3}(a1). In Fig.~\ref{fig:fig4} we show the effect of decompressing the systems shown in Figs.~\ref{fig:fig3}(a2-a4) back to $Z_s = 200$ after $t_w = 2^{18}\times 100$, and wait $2^{8} \times 100$ steps before measuring $P(\Delta_{ab})$. The difference between the decompressed systems and the reference one at large $\Delta_{ab}$ can be attributed to aging relaxation processes that have taken place in the stages presented in Figs.~\ref{fig:fig3}(a2-a4). They can be observed in Fig.~\ref{fig:fig4}, with an amplitude that decreases with increasing $Z$. 

These results suggest that a dynamic Gardner crossover can be defined from the distributions $P(\Delta_{ab})$ by focusing on the emergence of small values of $\Delta_{ab}$ upon compression, which is unrelated to aging relaxation dynamics. For such a criterion to coincide with the previously-used separation between the averages values $\langle \Delta_{ab}\rangle$ and $\langle \Delta \rangle$ which can be used for ultrastable glasses, we introduce the quantity
\begin{equation}
F(t_w) = \int_0^{\langle \Delta \rangle} P(\Delta_{ab}) \Delta_{ab},
\label{eq:F}
\end{equation}
which represents the enhanced probability that the distance between clones is smaller than the average size of the cage. For ultrastable glasses, $F(t_w)$ rapidly converges to a constant close to 0.5 for low pressures but fails to converge to 0.5 above the Gardner crossover (see Supplementary Fig.~\ref{app:app6}). For marginal and ordinary glasses, $F(t_w)$ decreases at low pressure due to aging relaxation but behaves similarly to ultrastable glasses at larger pressure. For all glasses, upon compressing the systems, $F(t_w)$ [displayed in Supplementary Fig.~\ref{app:app6}] changes from a decreasing function to an increasing one, indicating a generic dynamic crossover. We thus define a threshold pressure $Z_G$ at which $F(t_w)$ is nearly constant over long times. For the samples shown in Fig.~\ref{fig:fig3}, we can thus estimate the Gardner crossover to be $Z_G \approx 500$ for the ordinary glass, $Z_G \approx 400$ for the marginal glass, and $Z_G \approx 10^{3}$ for the ultrastable glass. The presence of subdiffusive motion\cite{hammond2020experimental} is not a unique signature of Gardner physics because it is also observed in glasses with simple aging dynamics.

The overall conclusion is that Gardner physics at large pressure is surprisingly robust against the aging relaxation dynamics taking place when the quasi-static conditions used in the theoretical analysis do not hold (See Supplementary Fig.~\ref{app:appunstable}). We confirm this conclusion by additional analysis of individual samples (See Supplementary Fig.~\ref{app:app4} and Fig.~\ref{app:app5}). Nevertheless, the complicated interplay between the various timescales involved in aging and Gardner dynamics certainly challenges analytic approaches to these non-equilibrium dynamics.

\subsection*{Increasing susceptibility upon approaching the jamming transition}

Having revealed that Gardner physics is pertinent in all preparation regimes, we quantify whether increasing collective fluctuations can also be detected. Following earlier work, we compute the global susceptibility $\chi_{AB}$ quantifying the fluctuations of the distance field between clones, $\Delta_{ab}(\vec{r})$ (see Methods). In the context of thermodynamic theory, the Gardner phase is critical and characterised by full replica symmetry breaking with an infinite correlation length and infinite susceptibility $\chi_{AB}$.  In the physical dimensions, simulations have shown that there exists a long but finite correlation length in the vicinity of the Gardner crossover, where $\chi_{AB}$ increases with pressure~\cite{LB18, BCJPSZ15,scalliet2019nature}, compatible with a possible divergence at finite pressure in $d=3$~\cite{BCJPSZ15,scalliet2019nature}.

However, most previous studies focused on ultrastable glass and the situation is unclear in the presence of relaxation dynamics at low pressure which can affect the behaviour of the susceptibility. To this end, we adopt a simpler measurement protocol where we do not change the pressure at the end of the preparation at $Z_s$ and directly create $N_c$ clones at $t_w \geq 0$ and $Z=Z_s$ which are then followed for $t_w>0$. We then monitor the time dependence of $\chi_{AB}(t_w)$ which quantifies collective effects. This procedure applies equally well to stable glasses (see Supplementary Fig.~\ref{app:appmerge}).

\begin{figure*}
\centering
\includegraphics[width=.95\linewidth]{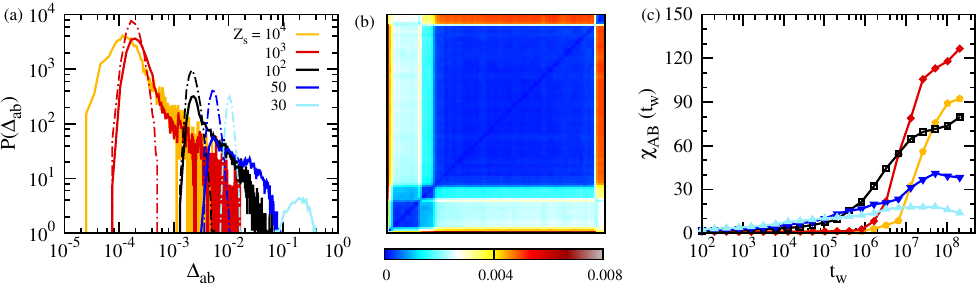}
\caption{Hierarchical landscape and increasing susceptibilities at larger pressures. (a) Probability distributions of mean-squared cage-relative distances between clones $P(\Delta_{ab})$ for $(\phi_g=0.600, t_s = 2^{18}\times 100)$, $(N_s = 100, N_c = 10)$ and various $Z=Z_s$. The solid curves are measured at long time, $t_w = 2^{18}\times 100$, the dotted lines at short times $t_w = 10^4-10^6$. (b) Representative heat map of $\Delta_{ab}$ for a sample with $(\varphi_g = 0.600, Z_s = 10^3, t_s = 2^{18}\times 100)$ using $N_c = 100$ clones at $Z = Z_s, t_w = 2^{18}\times 100$. The axes represent the clone index, and the value of $\Delta_{ab}$ is color coded. A clear hierarchical structure of phase space appears. (c) Time evolution of the susceptibility $\chi_{AB}(t_w)$ corresponding to the distributions shown in (a). Fast relaxation at small $Z$ does not result in the large susceptibility that Gardner physics yields at large pressure.}
\label{fig:fig5} 
\end{figure*}

We compare the probability distributions of clone distances at long times, $P(\Delta_{ab})$ for $t_w = 2^{18}\times 100$, with their short-time counterparts which contain only vibrations (measured using $t_w = 10^4 - 10^6$ depending on $Z_s$) in Fig.~\ref{fig:fig5}(a), using the initial states generated at various $Z_s$ for $(\varphi_g = 0.600, t_s = 2^{18}\times 100)$.

At short times, the distributions $P(\Delta_{ab})$ are Gaussian, centered at the averaged cage size for $Z_s = 30 - 10^3$, corresponding to the vibrations within the cages selected by the initial states. With increasing pressure, the time to reach a plateau in the MSD increases rapidly. At $Z_s=10^4$, the maximal time $t_w \sim 10^{8}$ is not long enough to see the MSD plateau. Therefore we do not display $P(\Delta_{ab})$ for this pressure in Fig.~\ref{fig:fig5}(a).

At long times, the evolution of the distributions $P(\Delta_{ab})$ demonstrates that clones gradually move away from each other and explore different parts of the landscape. At the lowest pressure $Z_s=30$, $t_w = 2^{18}\times 100$ is long enough for particles to escape from their cages, and $\Delta_{ab}$ is about one order of magnitude larger than the typical cage size. With increasing the pressure, $P(\Delta_{ab})$ remains much broader than the distribution of cage sizes, despite the slow dynamics. This suggests an increasing complexity of the landscape at higher pressures. Moreover, as shown by the heat map in Fig.~\ref{fig:fig5}(c) for a randomly chosen sample at $Z_s = 10^3$ and $t_w = 2^{18}\times 100$, the organisation of $\Delta_{ab}$ is reminiscent of the ultrametric structure characterising a Gardner phase, confirming the existence of a hierarchical landscape even in ordinary glasses which age.
  
Finally, we present in Fig.~\ref{fig:fig5}(c) the time evolution of the susceptibility $\chi_{AB}(t_w)$ for the same ordinary glass states as in Fig.~\ref{fig:fig5}(a). For $Z_s=30$, $50$, $\chi_{AB}(t_w)$ reaches a peak value $\chi^*_{AB}$ at $t_w = 10^{7}-10^{8}$, which corresponds to the timescale for escaping the initial cage. For higher pressures, $\chi_{AB}(t_w)$ remains small at short times but eventually increases at longer times, revealing increasingly long-ranged spatial correlations and stronger dynamics heterogeneities. In particular, $\chi_{AB}(t_w)$ at $Z_s = 10^3$ increases by about two orders of magnitude within the simulation time and continues to grow at large times. Interestingly, it can reach significantly larger values than the susceptibilities observed in marginal glasses~\cite{SZ18, LB18,scalliet2019nature} , because the dynamics is faster in ordinary glasses. We have also studied the dependence of $\chi_{AB}(t_w)$ on the packing fraction $\varphi_s$ by varying the preparation time $t_s$ at fixed $Z_s$ and find that $\chi^*_{AB}$ is larger for larger $\phi_s$ (see Supplementary Fig.~\ref{app:app8}). Therefore we conclude that a large susceptibility $\chi^*_{AB}$ emerges upon approaching jamming.

%\subsection*{Data Archival}
%PNAS must be able to archive the data essential to a published article. Where such archiving is not possible, deposition of data in public databases, such as GenBank, ArrayExpress, Protein Data Bank, Unidata, and others outlined in the \href{https://www.pnas.org/page/authors/journal-policies#xi}{Information for Authors}, is acceptable.

\section*{Discussion}

\begin{figure}
\centering
\includegraphics[width=.8\linewidth]{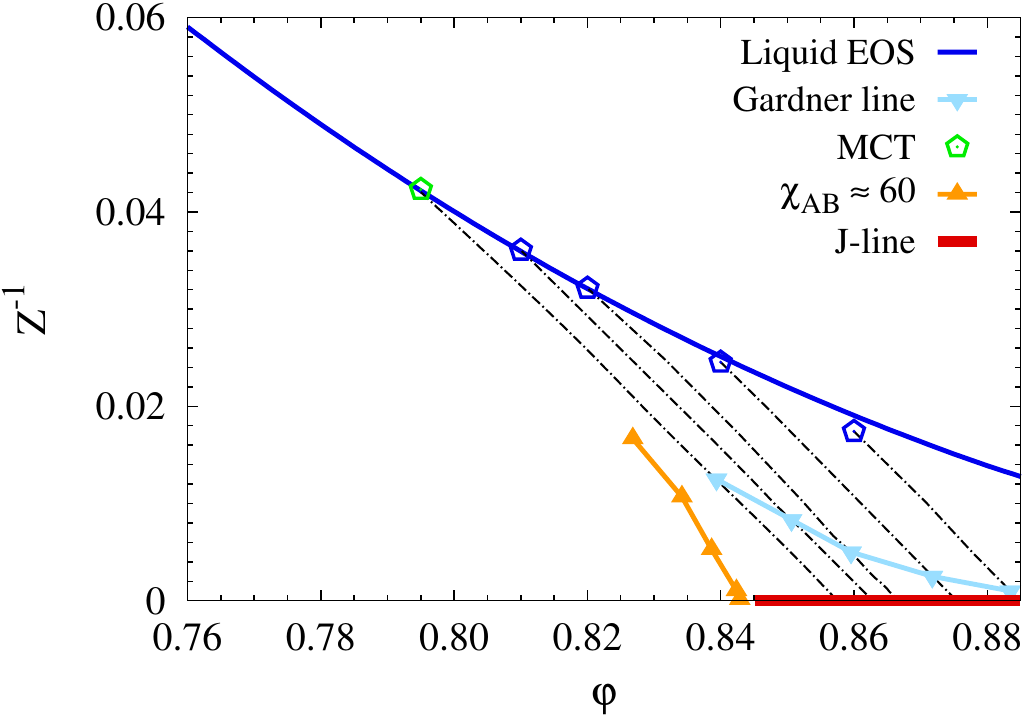}
\caption{Phase diagram of dense hard disks from the fluid to jamming. We summarise our results in the inverse pressure $1/Z$ versus packing fraction $\varphi$ representation. Equilibrium fluid states (hexagons) follow the Carnahan-Starling equation of state (blue line). The mode-coupling crossover is indicated by a green symbol. Black dotted lines represent glass equations of state obtained by compressions of stable equilibrium configurations starting from above $\phi_{MCT}$. A Gardner crossover separates simple and hierarchical glasses during these compressions (light blue symbols) before hitting the J-line at $Z \to \infty$ (red line). For ordinary glasses, there is no well-defined metastable state to follow but an isoline of susceptibility $\chi^*_{AB} \approx 60$ (orange symbols) locates the emergence of a complex landscape. Gardner physics is relatively pronounced in the triangular region `protecting' the J-line.} 
\label{fig:fig6}
\end{figure}

By using a broad variety of preparation histories leading to glasses with various degrees of stability, we have established that specific signatures due to the Gardner crossover are generically present in glasses compressed towards jamming. A careful choice of protocols and observables allowed us to reveal that the characteristics of a Gardner phase are systematically found in all types of glasses, and ultrastability is unnecessary to analyse the Gardner transition even though it makes the dynamics simpler to analyse. We summarise these observations and determine the complete region of the hard disk phase diagram where Gardner physics is pronounced in Fig.~\ref{fig:fig6}, which identifies the regions of state points where the glassy landscape becomes complex for a very broad range of initial conditions.

Hard spheres in large dimensions undergo a dynamic mode-coupling transition corresponding to the emergence of long-lived metastable states, but no such transition occurs in finite dimensions~\cite{CK93, gotze1999recent, parisi2020theory}. However, the dynamic crossover determined from an approximate power law fit inspired by mode-coupling theory to the equilibrium relaxation time remains a useful reference point. It is represented by a green symbol at ($\varphi_{MCT} \approx 0.795, Z_{MCT} \approx 24$) in Fig.~\ref{fig:fig6}.

For initial packing fractions $\varphi_g \geq \varphi_{MCT}$, we use the swap Monte Carlo algorithm to equilibrate the system. The measured equilibrium equation of state can be described by the Carnahan-Starling equation of state~\cite{mansoori1971equilibrium}. For these dense states, further compressions using ordinary MC simulations represent a relatively faithful implementation of the theoretical state-following construction, allowing us to measure the glass equations of states shown with black dotted lines in Fig.~\ref{fig:fig6}. Finally, using the observable $F(t_w)$ introduced in Eq.~(\ref{eq:F}), we locate the Gardner crossover for these stable glassy states, as depicted by the light blue line in Fig.~\ref{fig:fig6} which shows that $Z_{G}$ increases with $\varphi_g$. We find that $Z_G\approx 80$ for $\phi_g \approx \varphi_{MCT}$, which is significantly higher than the equilibrium value $Z_G \gg Z_{MCT} \approx 24$. In previous simulations of three-dimensional hard spheres~\cite{CKPUZ13, BCJPSZ15}, the authors did not consider the structural relaxation when determining the Gardner crossover in the vicinity of the mode-coupling crossover despite the lack of well-defined metastable states, similarly to our hard disk system. This might explain why conclude that the pressures $Z_G(\phi_{MCT})$ and $Z_{MCT}$ are distinct. This issue would be worth revisiting in three dimensions.

When $\varphi_g < \varphi_{MCT}$, there is no well-defined metastable basins in the equilibrium liquid phase. To evaluate the complexity of the free energy landscape, we measure the peak value of susceptibility over time $\chi^*_{AB}$ as in Fig.~\ref{fig:fig5}(c) which grows upon approaching jamming (occurring on the red line). To determine the relevance of Gardner physics, we empirically define the criterion $\chi^*_{AB} \approx 60$, which is large enough to signal the presence of Gardner physics, and low enough to be accessed within the simulation time window. The result is displayed with orange symbols in Fig.~\ref{fig:fig6}. This line does not exist below $Z \approx 60$ where the susceptibility remains small. The data for larger densities shown in Supplementary Fig.~\ref{app:app8} suggest that this line moves to larger pressures when $Z$ increases, see Fig.~\ref{fig:fig6}. 

Combining all measurements, we obtain a triangular zone in the phase diagram of Fig.~\ref{fig:fig6} where Gardner physics is highly relevant. Interestingly, this zone appears to `protect' the approach to the line of jamming points at $Z=\infty$. We conclude that when approaching the jamming transition using all kinds of the protocol by compressing ordinary, marginal, or ultrastable glasses, Gardner physics is inevitably involved. It is characterized by a hierarchically complex free energy landscape, leading to anomalous aging dynamics in the motion of dense assemblies of particles near the jamming transition, as well as specific physical properties related to marginal stability. Our simulations show that these signatures are robustly observed in two-dimensional systems, in the presence of thermal fluctuations, and for preparation protocols and compression histories similar to the ones studied in several experiments. Our study therefore naturally reconciles recent experimental and numerical studies of the Gardner crossover in dense particle systems, and provides a guide for future theoretical studies of their off-equilibrium dynamics in the vicinity of the jamming transition.

\matmethods{%Please describe your materials and methods here. This can be more than one paragraph, and may contain subsections and equations as required. 

\subsection*{Model} We simulate an assembly of $N = 1024$ two-dimensional hard disks in a squared box of length $L$ with periodic boundary conditions. The interaction between two disks is infinite when they overlap and zero otherwise. The disk diameter $\sigma$ is drawn randomly from a continuous distribution $P(\sigma) \propto \sigma^{-3}$ for $\sigma \in [0.45 \sigma_{max}, \sigma_{max}]$, and the polydispersity is $\sqrt{(\overline{\sigma^2}-\overline{\sigma}^2)}/\overline{\sigma} \approx 0.23$ with $\overline{\cdot \cdot \cdot}$ the average over $P(\sigma)$. The units for length, energy, and time are the mean diameter $\overline{\sigma}$, temperature $T$, and a MC sweep respectively. The physical control parameters are the volume fraction $\varphi = \pi\overline{\sigma^2}/4L^2$, the reduced pressure $Z = p/(\rho k_B T)$ and Monte Carlo (MC) sweep $t$, where $\rho = N/L^2$ is the particle number density. The equilibrium equation of state for this model can be fitted by the following empirical relation,
\begin{equation}
Z(\varphi) = \frac{1}{1-\varphi} + \frac{\overline{\sigma}^2}{\overline{\sigma^2}} \frac{(1+\varphi/8)\varphi}{(1-\varphi)^2}
\end{equation}

\subsection*{Observables} The reduced pressure can be computed by the contact number,
\begin{equation}
Z = 1 + \frac{1}{2N}\sum\limits^{N}_{i=1}\sum\limits^{N}_{j>i}\delta\left( \frac{R_{ij}}{\sigma_{ij}} - 1^{+} \right)
\end{equation}
where $\sigma_{ij} = (\sigma_i + \sigma_j)/2$ is the average diameter of particle $i$ and particle $j$, and $R_{ij} = | \vec{R}_i - \vec{R}_j |$ is the separation between two particles with $\vec{R}_i$ the coordinate of $i$. To get rid of the influence of the Mermin-Wagner fluctuations, we adopt cage-relative coordinates in our observations,
\begin{equation}
\vec{r}_i = \vec{R}_i - \frac{1}{N_i}\sum\limits^{N_i}_{j\in\partial i} \vec{R}_j
\end{equation}
where $\partial i$ is the set of neighbors of particle $i$ and $N_i = |\partial i|$. Here we define $j \in \partial i$ if $R_{ij} < 2\overline{\sigma}$.
In the state-following scheme, each of the $N_s$ samples is cloned $N_c$ times. We use the statistics of $(N_s = 100, N_c = 10)$ for general properties and $(N_s = 1, N_c = 100)$ when studying individual samples. The cage-relative distance between clones $a$ and $b$ is
\begin{equation}
\Delta_{ab} = \frac{1}{N} \sum\limits^N_{i=1} \Delta_{ab, i}, \quad \quad \Delta_{ab, i} = |\vec{r}^a_i - \vec{r}^b_i |^2
\end{equation}
Averaging over the clone pairs and over samples, one can get the mean-squared clone distance,
\begin{equation}
\langle \Delta_{ab} \rangle = \frac{1}{N_s}\sum\limits^{N_s}_{\alpha=1}\left[ \frac{2}{N_c (N_c -1)} \sum\limits^{N_c}_{a, b\in\partial \alpha, b>a} \Delta_{ab} \right]
\end{equation}
Here $\partial \alpha$ is the set of clones of a sample $\alpha$, $\langle \cdot \cdot \cdot \rangle$ represent averaging over samples and clone pairs.
The susceptibility for the spatial displacement field is,
\begin{equation}
\chi_{AB} = N \frac{\langle \Delta^2_{ab} \rangle - \langle \Delta_{ab} \rangle^2 }{\overline{\langle \Delta^2_{ab, i} \rangle - \langle \Delta_{ab, i} \rangle^2 }} 
\end{equation}

We probe the dynamics with two-time mean-squared displacement (MSD),
\begin{align*}
\Delta(t_w, t) = \frac{1}{N_s}\sum\limits^{N_s}_{\alpha=1}\left[ \frac{1}{N_c} \sum\limits^{N_c}_{a\in\partial \alpha} \Delta_a(t_w, t) \right], \\
\Delta_a(t_w, t) = \frac{1}{N}\sum\limits^{N}_{i=1}|\vec{r}^a_i(t_w + t) - \vec{r}^a_i(t_w) |^2
\end{align*}
with $t_w$ the initial time for the observation and $t$ the time window.
}

\showmatmethods{} % Display the Materials and Methods section

\acknow{We thank M. Ozawa and F. Zamponi for helpful discussions. The research has been supported by the National Natural Science Foundation of China Grants (No. 11734014, No. 11975295, and No. 12047503) and the Chinese Academy of Sciences Grants(No. QYZDJ-SSW-SYS018 and No. XDPD15). L.B. acknowledges support from the Simons Foundation (No. 454933 L. B.). We also thank the Supercomputing Center of University of Science and Technology of China for the computer time. }

\showacknow{} % Display the acknowledgments section

% Bibliography
\bibliography{pnas-sample}

%\clearpage

%\begin{appendices}
\section*{\large \bf Supplementary Information} 

\subsection*{\bf A. State-following analysis for ultrastable glasses}

\begin{figure}
\centering
\includegraphics[width=.8\linewidth]{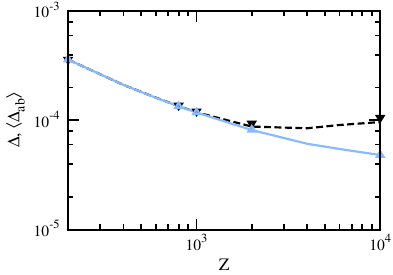}
\caption{The long-time cage-relative MSD $\Delta$ for $t_w = t = 2^{17}\times 100$ (blue) and the cage-relative mean-squared clone distances $\langle \Delta_{ab} \rangle(t_w = 2^{17}\times 100)$ (black) as a function of pressure. The points are measured from the protocols described in the main for ultrastable glasses with $(\varphi_g = 0.860, Z_s = 200, t_s = 2^{18}\times 100)$, and the lines are the corresponding data for state-following equilibrium glasses with $\varphi_g = 0.860$.}
\label{app:appultra}
\end{figure}

We compare the results using the state-following protocol used in earlier work to the results in the main text for ultrastable glass states. Specifically the state-following protocol is used from the equilibrium states of $\varphi_g = 0.860$, and it is compared to the ultrastable states prepared at $(\varphi_g = 0.860, Z_s = 200, t_s = 2^{18}\times 100)$ used in the main text. As shown in Fig.~\ref{app:appultra}, the two protocols give consistent results. In both cases, the long-time limits of cage-relative MSD $\Delta$ and mean-squared clone distances $\langle\Delta_{ab}\rangle$ are roughly equal when $Z < Z_{G} \approx 10^3$ and deviate from one another upon increasing the pressure further.

Therefore, the Gardner crossover using the protocols of ultrastable glass samples agrees well with the result obtained by the state-following protocol for corresponding metastable equilibrium glasses.

\subsection*{\bf B. Complex dynamics at high pressures}

\begin{figure*}
\centering
\includegraphics[width=.95\linewidth]{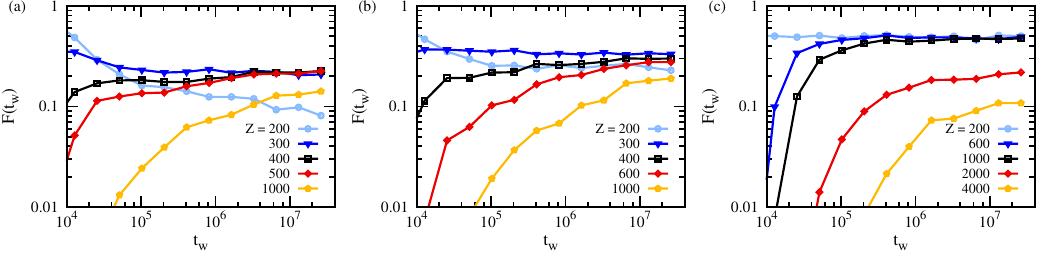}
\caption{Probability that the clone distance is smaller than the mean-squared cage size as the function of waiting time. The protocols are the same as in Fig.~\ref{fig:fig2} and Fig.~\ref{fig:fig3} where the systems are prepared by compressing equilibrium states at $\varphi_g = 0.600$ in (a), $0.820$ in (b) and $0.860$ in (c) to $Z_s = 200$ for a time $t_s = 2^{18}\times 100)$, resulting in ordinary, marginal and ultrastable glasses. In all cases, $F$ increases with $t_w$ at high enough pressures, corresponding to falling into sub-basins. The crossovers to these dynamics are estimated at $Z_G \approx 500$, $400$, and $10^3$.}
\label{app:app6}
\end{figure*}

To better characterise the crossover to the complex dynamics in arbitrary protocols, we compute the evolution of the probability for the clone distance to be smaller than the typical cage size,
\begin{equation}
F(\Delta_{ab} \le \langle \Delta \rangle) = \int^{\langle \Delta \rangle}_{0} d\Delta_{ab} \ P(\Delta_{ab}).
\end{equation}
Here the mean-squared cage size $\langle \Delta \rangle$ can be measured from the plateau value of $\Delta (t_w, t)$, and it captures the averaged size of the basins at a given $Z$ condition, and $P(\Delta_{ab})$ is the probability distribution function of clone mean-squared distances. For simple glasses in the absence of aging relaxation, $P(\Delta_{ab})$ is approximately Gaussian with the mean value $\Delta_{ab} \approx \langle \Delta \rangle$, corresponding to $F \approx 0.5$, as shown by the results for ultrastable glasses at $Z = 200$ and $Z = 600$ in Fig.~\ref{app:app6}(c). On the other hand, if the dynamics is dominated by aging diffusion dynamics, most of the clones jump out of the initial basin and explore different portions of the landscape. As a result, $F(t_w)$ decreases to a smaller value as more clone distances become larger. This occurs when following ordinary and marginal glasses at relatively low pressures, see Figs.~\ref{app:app6}(a, b). Upon compression, all three glasses to high enough pressures, the dynamics becomes slow, accompanying the emergence of a complex landscape, leading to the overall decrease of $F(t_w)$, as shown in Fig.~\ref{app:app6}. Meanwhile, as some clones cross some barriers and fall into the same more stable sub-basins, $F(t_w)$ grows for a long time. As shown by the results at high pressures in Fig.~\ref{app:app6}, the clone dynamics of ordinary glasses and marginal glasses is analogous to the case of ultrastable glasses, and can thus be used to determine a dynamic Gardner crossover. 

\subsection*{\bf C. Robustness of Gardner physics}

\begin{figure*}
\centering
\includegraphics[width=.75\linewidth]{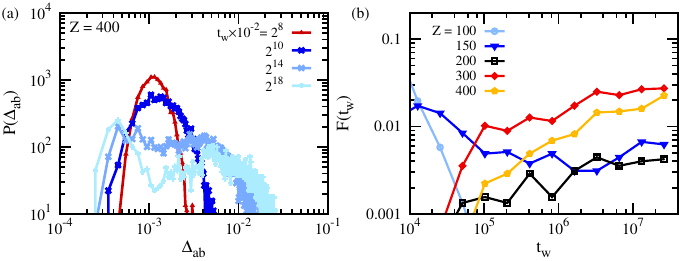}
\caption{(a) Evolution of the distributions of mean-squared cage-relative distances between clones $P(\Delta_{ab})$ at $Z = 400$ for the less stable samples compressed from the equilibrium states of $\phi_g = 0.600$ on the condition of $(Z_s = 10^{2}, t_s = 2^{10}\times 100$) shown in Fig.~\ref{fig:fig1}. (b) The cumulative probability for the clone distance being smaller than the cage size as the function of waiting time. The states are respectively generated at $Z = 100, 150, 200, 300$ and $400$ using the samples. The crossover is estimated at $Z_G\approx 200$.}
\label{app:appunstable}
\end{figure*}

We study the dynamic Gardner crossover by compressing the ordinary glasses produced at shorter $t_s$. As an example, we show the result of the glass prepared using the parameters of $\varphi_g = 0.600, Z_s = 100$ and $t_s = 2^{10}\times 100$. As shown in Fig.~\ref{fig:fig2}(b), the dynamics of representative glass exhibits strong structural relaxations, and the plateau of the MSD $\Delta(t)$ is nearly observed indicating the formation of transient cages. We employ these highly unstable glasses as our samples and compress them at various pressures as in the main text. In Fig.~\ref{app:appunstable}(a), we plot the evolution of the probability distribution of mean-squared clone distances $P(\Delta_{ab})$ at $Z = 400$, from which one can see that $P(\Delta_{ab})$ develop a peak centered at $\langle\Delta\rangle$ with time. It suggests a dynamic Gardner crossover for this protocol. We then compute the cumulative probability $F(t_w)$ to determine threshold pressure. As shown by Fig.~\ref{app:appunstable}, when increasing $Z$ larger than $Z_G \approx 200$, $F(t_w)$ turns to increase over time which is qualitatively similar to the results of more stable glasses in Fig.~\ref{app:app6}. We thus conclude that Gardner physics is considerably robust against structural relaxation in hard disk glasses.

\subsection*{\bf D. Evidence for hierarchical dynamics}

\begin{figure*}
\centering
\includegraphics[width=.95\linewidth]{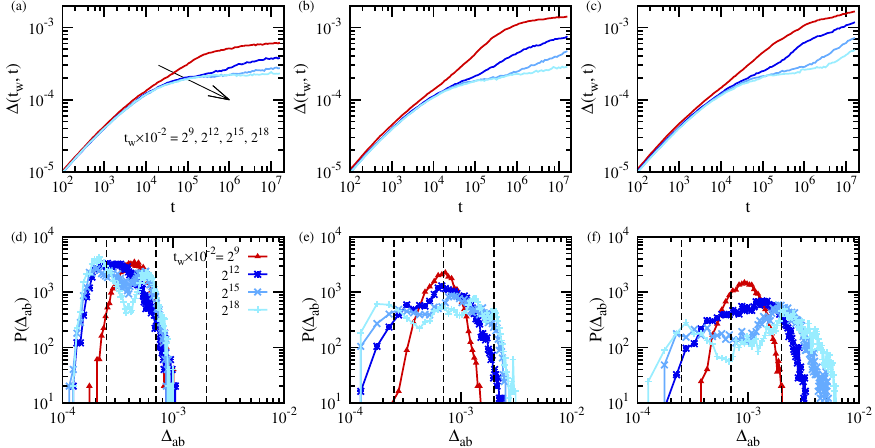}
\caption{(a-c) The cage-relative MSDs $\Delta(t_w, t)$s for the protocols of following three individual ordinary glass samples of $(\varphi_g = 0.600, Z_s = 100, t_s = 2^{18}\times 100)$ at a intermediate pressure higher than the crossover pressure $Z = 800 > Z_G \approx 300$. $N_c = 100$ clones are used for each of the three samples. (d-f) The related probability distributions of clone displacement $P(\Delta_{ab})$s. Here the results shown in the same column correspond to the same protocol. The values of mean-squared cage sizes $\Delta_{EA}(Z = 800) \approx 2.5\times 10^{-4}$, $\Delta_{EA}(Z = 300)\approx 7\times 10^{-4}$ and $\Delta_{EA}(Z = 100)\approx 2\times 10^{-3}$ are denoted with the vertical dashed lines.}
\label{app:app4} 
\end{figure*}

In this section, we analyze the correspondence between sampling the glass landscape and aging dynamics for ordinary glasses by focusing on individual samples. We study representative samples prepared from compressing independent configurations at $\phi_g = 0.600$ to $Z_s=100$ for $t_s = 2^{18}\times 100$, using $N_c = 100$ clones for each sample. The dynamic Gardner crossover for such ordinary glasses is estimated at $Z_{G} \approx 300$. For the system size $N = 1024$, we observe strong sample-to-sample fluctuations. The aging behaviors are roughly classified into three categories within the time window $\sim 3 \times 10^{7}$, as illustrated in Fig.~\ref{app:app4}. The overall amplitude of the aging dynamics varies from one sample to another, corresponding to distinct evolution of the distribution of clone displacement $P(\Delta_{ab})$.

To better reveal the hierarchical structure of the landscape, we indicate the three characteristic mean-squared cage sizes $\langle\Delta\rangle$ for the protocols at $Z = Z_s = 100$, $Z = Z_G = 300$, and $Z = 800$. As seen in Fig.~\ref{app:app4}(d), $P(\Delta_{ab})$ for the first sample develops a bimodal structure with increasing $t_w$, with two peaks centered at $\langle \Delta \rangle(Z = 800)$ and $\langle \Delta \rangle (Z = 300)$. This suggests the emergency of smaller clusters of clones, where the typical distance between within each cluster is $\langle \Delta \rangle(Z = 800)$, while the average distance between clusters is $\langle \Delta \rangle (Z = 300)$.

As shown in Fig.~\ref{app:app4}(b), $P(\Delta_{ab})$ for the second sample develops a tail up to $\langle \Delta \rangle (Z = 100)$, meaning that the whole basin selected by the initial state at $Z_s = 100$ remains dynamically accessible. Correspondingly, this sample shows stronger aging dynamics, compare to Figs.~\ref{app:app4}(a, b).

The third sample, shown in Figs.~\ref{app:app4}(c, f) exhibits the strongest aging dynamics among the three cases, and the clone distances can significantly exceed $\langle \Delta \rangle (Z = 10^{2})$, implying the existence of structural relaxation beyond the initial basin. Note that this classification is empirical since the simulation duration is limited. When $t_w$ is long enough, all samples should resemble the third example. Nevertheless, there are growing peaks at the small distances near $\langle \Delta \rangle (Z = 800)$ at long times in all cases, indicating that some clones remain close to one another, which is indicative of Gardner physics.

\begin{figure*}
\centering
\includegraphics[width=.95\linewidth]{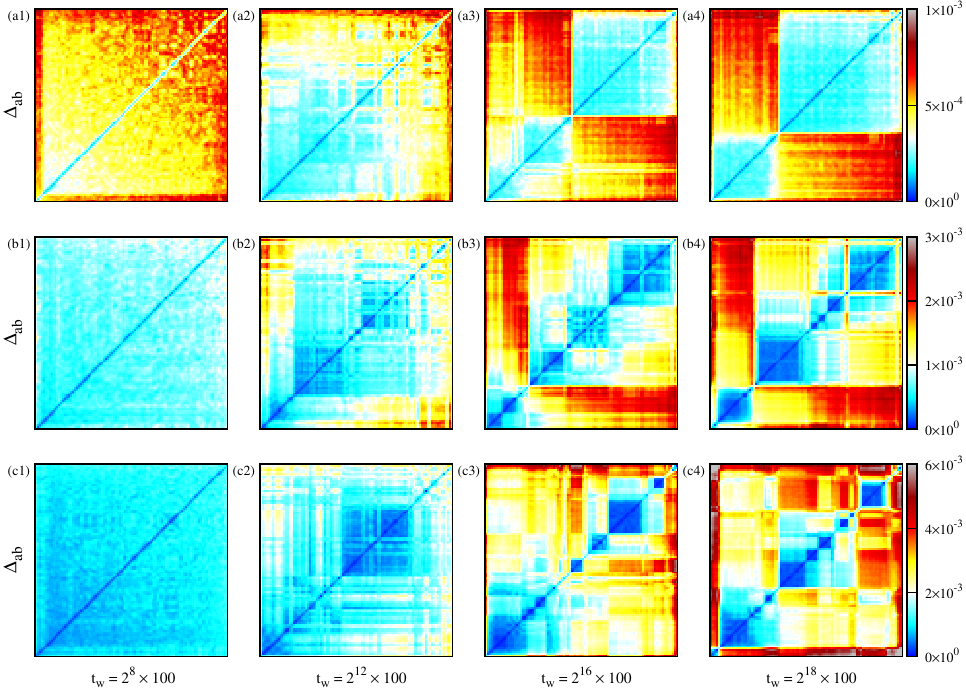}
\caption{Heat maps constructed with $N_c = 100$ clones for $\Delta_{ab}$ between clones copied from different ordinary glass samples taken at $(\varphi_g = 0.600, Z_s = 10^{2}, t_s = 2^{18}\times 100)$ are shown in different rows. Here the following pressure is $Z = 800$ as in Fig.~\ref{app:app4}. For each column, the waiting time after crunches is fixed and varies from $2^{8} \times 100$ to $2^{18} \times 100$ as indicated by labels. For each sub-graph, the axes and the color represent the clone indexes and $\Delta_{ab}$ respectively.}
\label{app:app5} 
\end{figure*}

We next visualise the topological structure of the landscape detected by the clones for the three samples at $Z = 800$. We plot the heat maps of clone distances $\Delta_{ab}$ for the states in Fig.~\ref{app:app4} in Fig.~\ref{app:app5}. The choice of clone ordering is irrelevant to the structure of the data and is only useful for illustrative purposes. The current choice is based the clustering order. In the first sample, two major clusters form at intermediate time scales, and one of them shrinks at longer $t_w$ as clones eventually end in the more stable sub-basin. This is similar to the situation seen when compressing the ultrastable glasses at $Z > Z_G$.

By contrast, the situation for the second and third samples is more complicated, since the time scales of aging events in different glass basins are close. The heatmap reveals the presence of independent basins (explored via aging relaxation) in which a hierarchical Gardner structure emerges, see Figs.~\ref{app:app5}(b, c). These hierarchical heat maps reveal the intricate structure underlying the broad distributions in Figs.~\ref{app:app4}(e, f). 

\subsection*{\bf E. Susceptibility analysis for stable states}

We consider the glass states prepared by compressing the equilibrium glasses of $\varphi_g = 0.820 > \varphi_{MCT} \approx 0.795$ at $t_s = 2^{18} \times 100$ and different $Z_s = 10^5$, $10^4$, $10^3$, $300$ and $10^2$. The Gardner crossover is found to occur near $Z_{G}(\varphi_g = 0.820) \approx 200$ when using the standard state-following scheme. Thus, the glasses produced at $Z_s \geq 200$ are marginally stable.

\begin{figure*}
\centering
\includegraphics[width=.95\linewidth]{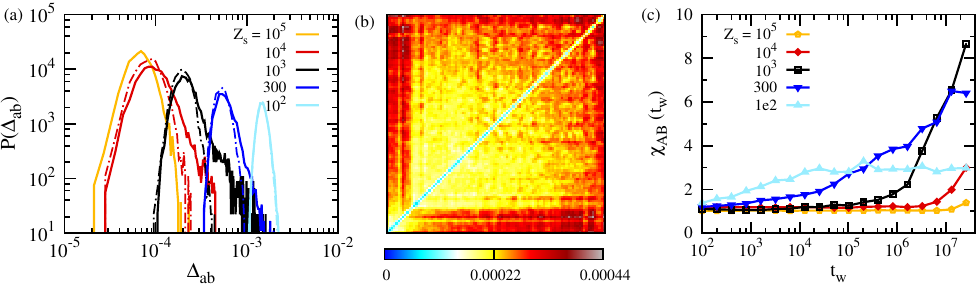}
\caption{(a) Probability distributions of mean-squared clone distances $P(\Delta_{ab})$ for the protocol of various $Z=Z_s$ using $t_s = 2^{18}\times 100$ and $(N_s = 100, N_c = 10)$ at $t_w = 2^{18} \times 100$. From left to right, the pressures are $Z = Z_s = 10^5$, $10^4$, $10^3$, $300$ and $10^2$. The short-time vibration behaviors are conveyed by dotted curves acting as references.  At $Z_s = 10^5$, $t_w = 2^{18}\times 100$ is not long enough to form the well-defined cages. We thus are not able to present the Gaussian-like reference for $Z_s = 10^5$. (b) Heat map for a representative sample with $(\varphi_g = 0.820, Z_s = 10^3, t_s = 2^{18}\times 100)$ at $Z = Z_s$ and $t_w = 2^{18}\times 100$. (c) Evolution of the corresponding susceptibility $\chi_{AB}(t_w)$ for the same protocols in (a).}
\label{app:appmerge}
\end{figure*}

When we make clones at the pressure of $Z = Z_s$, as in the main text, the corresponding susceptibility has the behaviour shown in Fig.~\ref{app:appmerge}(c). For $Z = 10^{2} < Z_{G}$, $\chi_{AB}(t_w)$ reaches a small value $\chi^*_{AB}\sim 3$ after a time scale $t_w \sim 10^{4}$. By contrast, the susceptibility for marginal glasses of $Z \geq Z_{G}$ grows over the entire time window, owing to the slow dynamics in the presence of small barriers. We conclude that the susceptibility analysis presented in the main text provides results consistent with the state-following analysis used in previous studies when the latter can be used. In Fig.~\ref{app:appmerge}(a), we report the long-time distributions $P(\Delta_{ab})$ for the same data. In line with the behavior of $\chi_{AB}$, $P(\Delta_{ab})$ is Gaussian at $Z = 10^{2}$. Upon compressing to $Z \geq Z_G$, there are activated events in the Gardner region. As a result, $P(\Delta_{ab})$ extends to larger $\Delta_{ab}$ with respect to the Gaussian distributions of short time vibrations. We also display the organization of the clones for $(Z = 10^3, t_w = 2^{18}\times 100)$ for a representative sample of $(Z_s = 10^{3})$ in Fig.~\ref{app:appmerge}(b).

In summary, the analysis of marginal glasses following the analysis proposed in the main text leads to qualitatively similar findings as for ordinary glasses, even though the structural relaxation is highly suppressed for the marginal states.

\subsection*{\bf F. Susceptibilities upon approaching jamming}

\begin{figure}
\centering
\includegraphics[width=.8\linewidth]{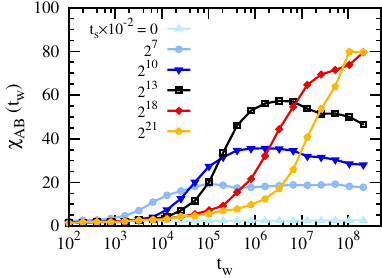}
\caption{Evolution of susceptibility $\chi_{AB}(t_w)$ for $(\varphi_g = 0.600, Z = Z_s = 10^2)$ at different preparation times $t_s$. The corresponding packing fractions are $\varphi_s \approx 0.8081, 0.8320, 0.8346, 0.8355$ and $0.8361$ at $t_s = 2^{7}\times 100, 2^{10}\times 100, 2^{13}\times 100, 2^{18}\times 100$ and $2^{21}\times 100$ respectively. With the densification at longer $t_s$, ordinary glasses become more stable and show larger $\chi^{*}_{AB}$.}
\label{app:app8}
\end{figure}

In the main text, we show that the susceptibility $\chi_{AB}$ extracted from the evolution of ordinary glasses at $Z = Z_s$ grows with increasing pressure. Here we report the dependence of $\chi_{AB}(t_w)$  on the density for ordinary glasses prepared at fixed $(\varphi_g = 0.600, Z_s = 10^2)$. When rapidly compressing these fluid states to $Z_s > Z_{MCT}$, the system falls out of equilibrium and undergoes evident densification at long times, as shown in Fig.~\ref{fig:fig1}. Hence $t_s$ controls the density of ordinary glasses. In Fig.~\ref{app:app8}, we show the evolution of $\chi_{AB}(t_w)$ for the initial states generated at $t_s = 0 - 2^{21}\times 100$. One can see that $\chi_{AB}(t_w)$ takes small values for short $t_s$. With increasing the time $t_s$, the stability, and density increase, the structural relaxations are less pronounced, and the long-time susceptibility becomes larger. 

%\end{appendices}

\end{document}